\documentclass[aps,twocolumn,showpacs,superscriptaddress,groupedaddress,floatfix, twocolumn,aps,prd,amsmath,amssymb]{revtex4}
\usepackage[english]{babel}
\usepackage[utf8]{inputenc}
\usepackage{epsfig}
\usepackage{multirow}
\usepackage{graphicx}
\usepackage{graphics}
\usepackage{subfigure}
\usepackage{epstopdf}
\usepackage{float}
\usepackage{bm}
\usepackage{braket}
\usepackage{lineno}
\usepackage{slashed}
\usepackage{hyperref}
\hypersetup{colorlinks=true, linkcolor=red, filecolor=black, citecolor=blue}
\newcommand{\PR}[1]{\ensuremath{\left[#1\right]}}
\newcommand{\PC}[1]{\ensuremath{\left(#1\right)}}

\begin{document}

\title{Effects of the two-dimensional Coulomb interaction in both Fermi velocity and energy gap for Dirac-like electrons at finite temperature}

\author{Nilberto Bezerra}
\email{jose.bezerra@icen.ufpa.br}
\affiliation{Faculdade de Física, Universidade Federal do Pará, Avenida Augusto Correa 01, 66075-110, Belém, Pará,  Brazil}

\author{Van Sérgio Alves}
\email{vansergi@ufpa.br}
\affiliation{Faculdade de Física, Universidade Federal do Pará, Avenida Augusto Correa 01, 66075-110, Belém, Pará,  Brazil}

\author{Leandro O. Nascimento}
\email{lon@ufpa.br}
\affiliation{Faculdade de Física, Universidade Federal do Pará, Avenida Augusto Correa 01, 66075-110, Belém, Pará,  Brazil}
\affiliation{Universidade Federal de Campina Grande, Rua Aprígio Veloso 882, 58429-900, Campina Grande, Paraíba, Brazil}

\author{Luis Fernández}
\email{luis.fernandez@ufrontera.cl}
\affiliation{Departamento de Ciencias Físicas, Facultad de Ingeniería y Ciencias, Universidad de La Frontera, Avenida Francisco Salazar 01145, Casilla 54-D, Temuco, Chile}

\date{\today}

\begin{abstract}
We describe both the Fermi velocity and the mass renormalization due to the two-dimensional Coulomb interaction in the presence of a thermal bath. To achieve this, we consider an anisotropic version of pseudo quantum electrodynamics (PQED), within a perturbative approach in the fine-structure constant $\alpha$. Thereafter, we use the so-called imaginary-time formalism for including the thermal bath. In the limit $T\rightarrow 0$, we calculate the renormalized mass $m^R(p)$ and compare this result with the experimental findings for the energy band gap in monolayers of transition metal dichalcogenides, namely, WSe$_2$ and MoS$_2$. In these materials, the quasi-particle excitations behave as a massive Dirac-like particles in the low-energy limit, hence, its mass is related to the energy band gap of the material. In the low-temperature limit $T\ll v_F p $, where $v_F p$ is taken as the Fermi energy, we show that $m^R(p)$ decreases linearly on the temperature, i.e, $m^R(p,T)-m^R(p,T\rightarrow 0)\approx -A_\alpha T +O(T^3)$, where $A_\alpha$ is a positive constant. On the other hand, for the renormalized Fermi velocity, we find that $v^R_F(p,T)-v^R_F(p,T\rightarrow 0)\approx -B_\alpha T^3 +O(T^5)$, where $B_\alpha$ is a positive constant. We also perform numerical tests which confirm our analytical results.

\end{abstract}

\maketitle

\section*{INTRODUCTION}

The study of field theories in (2+1) dimensions has led to important insights and applications in a variety of fields, including high-energy physics \cite{Burden, Maris} and condensed matter physics \cite{Prange}. In the former the idea is to obtain a simplified version of quantum chromodynamics that yields both confinement and chiral symmetry breaking \cite{Maris}. In the latter, after the experimental realization of two-dimensional materials, the goal is to derive field theories for describing the electron-electron interaction, which is sometimes neglected in a simplified condensed-matter-physics model \cite{MarinoBook, Boulevard}. Two-dimensional materials, such as graphene, possess unique electronic properties that may be explained within a (2+1) dimensional field theory. In particular, as an effect of the honeycomb lattice, its quasi-particles obey a Dirac-like equation with an effective Fermi velocity $v_F\approx c/300$ and an effective mass $m$, which for graphene vanishes while for monolayers of transition metal dichalcogenides (TMDs) is in the order of 100 meV. This relativistic-like equation for the matter field allow us to include the electromagnetic interaction in terms of a gauge theory, similar to what is done in quantum electrodynamics in (3+1) dimensions. Obviously, such approach is expected to work well only for a low-energy description, which is enough for explaining transport properties \cite{Neto2009, PRX}, renormalized parameters \cite{Luis2020, Vozmediano1994}, and anomalies \cite{PRX, Luis2020}.

A field theory model that describes the true electromagnetic interaction in these two-dimensional materials must take into account that electrons live in two spatial dimensions while photons live in three dimensions. The construction of such model was proposed in Ref.~\cite{Marino1993} and is referred to as pseudo quantum electrodynamics (PQED) due to the presence of a pseudo-differential operator \cite{Amaral}. This theory has been proven to be unitary \cite{Marino2014}, causal \cite{Amaral}, and conformal invariant \cite{Ana, Amos, MarinoBook}. Furthermore, it has been successfully used to explain some properties in graphene and TMDs. For instance, considering graphene, it has provided a theoretical description for the quantum valley Hall effect, quantum corrections for the dc longitudinal conductivity \cite{PRX}, and the electron g-factor \cite{FatorG}. For TMDs, PQED has been used to describe the excitonic spectrum and the renormalization of the band gap at zero temperature \cite{Excitons2018, Luis2020}. Other theoretical results concern the chiral symmetry breaking in both zero \cite{Alves2013} and finite temperature \cite{Leandro, Luis2021} as well as the inclusion of an external magnetic field in Ref. \cite{Menezes2016}. The effect of a grounded conducting surface and a cavity in the vicinity of a graphene sheet also have been considered in Refs.~\cite{Danilo2017,Cavity}, respectively.  It is worth to mention that PQED is also known as reduced quantum electrodynamics \cite{Miransky} and several results have been related to this theory \cite{Teber, Ana, Amos, Pedrelli}. As it may be concluded from this short list, the effects of finite temperatures are less discussed in the literature. 

A full description of thermal fluctuations in a many-particle quantum system made of electrons is a very hard task, even when considering independent electrons. Usually one assumes a low-temperature regime, where the Fermi-Dirac distribution becomes a step function, implying that all states below the Fermi energy are ocuppied and all states above it are empty. Although this is not a general solution, it has some important aplications in the description of transport properties in a metal \cite{Ashcroft}. Notice that the contribution of thermal effects to the renormalization of the energy band gap in TMDs and to the Fermi velocity have not been considered yet. 

In this paper, we describe the influence of the electron-electron interaction in both $m$ and $v_F$, using PQED at finite temperature. For $T = 0$, we compare our results with the experimental findings for the energy band gap, for tungsten diselenide (WSe$_2$) \cite{Fig4_WSe2} and molybdenum disulfide (MoS$_2$) \cite{Fig4_MoS2}, and a good agreement is found. Surprisingly, our main results are consistent with the RPA (random phase approximation) and large-$N$ expansion \cite{Luis2020, DasSarma2014}, where the coupling constant may be large. On the other hand, the results for $v_F$ have been discussed in Ref.~\cite{Vozmediano1994}. For $T \neq 0$, we apply the Matsubara formalism, also know as the imaginary-time formalism \cite{Matsubara}. Through the electron self-energy, we derive analytical expressions for the renormalized parameters. Thereafter, these results are confirmed by numerical solutions of the renormalized parameters for any temperature.

This work is organized as follows. In Sec.~\ref{The_Model}, we present the model and its Feynman rules. 
In Sec.~\ref{Sec_R_without_T}, we calculate the electron self-energy at $T=0$ and obtain the renormalization of both the Fermi velocity and the mass. In Sec.~\ref{The_self-energy_at_T}, we use the Matsubara formalism to introduce the temperature in our theory in order to calculate the electron self-energy at $T \neq 0$. In Sec.~\ref{The_Renormalization_at_T}, we investigate the influence of finite temperature in both the renormalization of the Fermi velocity and  the mass. In Sec.~\ref{Discussions}, we review and discuss our main results. We also include two Appendixes, where some details of the calculations are provided.

\section{The model}\label{The_Model}
Let us start by considering the anisotropic version of the PQED theory, given by
\begin{equation}\label{L_isotropic}
\begin{split}
	\mathcal{L} =& \frac{1}{4} F_{\mu\nu}\frac{2}{\sqrt{-\Box}}F^{\mu\nu} + \bar{\psi}_a (i \gamma^{0}\partial_{0} +i v_{F}\gamma^{i} \partial_{i} - m)\psi_a  \\
	&+ e\,\bar{\psi}_a\left( \gamma^{0}A_{0} +v_{F}\gamma^{i} A_{i} \right) \psi_a  - \frac{\xi}{2} A_{\mu} \frac{\partial^{\mu}\partial^{\nu}}{\sqrt{-\Box}}A_{\nu} , 
\end{split}
\end{equation}
where $ F^{\mu\nu} $ is the usual field-intensity tensor of the U(1) gauge field $A_\mu$ and the parameter $\xi$ is the gauge-fixing term. ${\psi_a}$ is the Dirac field describing the electrons in a two-dimensional material, whose representation is  $ \psi^\dagger_a = (\psi^{\star}_{A\uparrow}\,\psi^{\star}_{A\downarrow}\,\psi^{\star}_{B\uparrow}\, \psi^{\star}_{B\downarrow})$, a four-component Dirac spinor with sublattices $A$ and $B$ as well as with spin orientataions $\uparrow$ and $\downarrow$. Furthermore, $m$ is a mass term for the matter field that describes a possible energy gap in the Dirac points. The $\gamma^\mu$ are rank-4 Dirac matrices that obey an anti-commutator relation, given by $\{\gamma^\mu, \gamma^\nu \} = -2 \delta^{\mu\nu} $ in the Euclidean space-time. Finally, $\Box \equiv \,\Delta-\partial^2/\partial t^2$ is the D'Alembertian operator and $v_F$ is the Fermi velocity of electrons in a two-dimensional material. Let us use the following convention $c=\hbar=k_B=1$.

The usual Feynman rules for the model in Eq.~(\ref{L_isotropic}) are 
\begin{equation}\label{ElectronPropagator}
	S_{0F} (\bar p) = \frac{-1}{\gamma^\mu \bar p_\mu - m}, 
\end{equation}
describing the fermion propagator, where ${\bar p}^{\mu}=(p_0,v_F \textbf{p})$ and $\bar p^2=p_0^2+v_F^2 \textbf{p}^2$. Note that the pole of the fermion propagator provides the energy dispersion, given by $p_0 = E(\textbf{p})=\pm  \sqrt{v_F^2 \textbf{p}^2+m^2}$. When considering the comparison with two-dimensional materials, $2m$ is equal to the energy gap for the system. For graphene, $m=0$ it reproduces the tight-binding result for the electron \cite{Neto2009}. On the other hand, the gauge-field propagator reads
\begin{equation}\label{GaugeField}
	\Delta_{\mu\nu} ( p) = \frac{1}{2\varepsilon\sqrt{ p^2}}\PR{g_{\mu\nu}-\PC{1-\frac{1}{\xi}}\frac{ p_{\mu} p_{\nu}}{ p^2}},
\end{equation}
where $ p_{\mu}$ is the three-momentum, given by $ p_{\mu}=(p_0, \textbf{p})$ with $ p^2=p_0^2+ \textbf{p}^2$ and $\varepsilon$ is a constant that describes the dielectric constant. Finally, 
\begin{equation}
	\Gamma^{\mu}= e\PC{ \gamma^{0}, v_F\,\gamma^{j}}
\end{equation} 
is the interaction vertex.

It is well known that to obtain quantum corrections to both $v_F$ and $m$, we must calculate the electron self-energy. Although this has been done in Refs.~\cite{Vozmediano1994, Luis2020} at zero temperature, the effects of a thermal bath have been not discussed until now. Here, we shall consider these effects and include its contribution by using the so-called imaginary-time formalism, which introduces the Matsubara frequencies in the bare propagators of the theory \cite{DasBookThermo}.

The model in Eq.~\eqref{L_isotropic} describes the full electromagnetic interaction for two-dimension Dirac-like electron. This allows one to calculate dynamical effects, such as anomalies and chiral symmetry breaking \cite{PRX, Luis2021}. Here, we shall consider a simplified version of this model that applies when the Fermi velocity is much less than the light speed. This is called the static limit.

\begin{figure}[h!]
	\centering
	\includegraphics[width=0.4\textwidth]{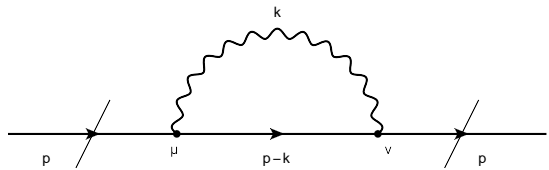}
	\caption{ The electron self-energy. The straight line represents the fermion propagator while the waved line denotes the gauge-field propagator.}\label{Fig_SelfEnergy}
\end{figure}
In the static limit the vertex interaction is taken as $\Gamma^\mu \to e \gamma^0$ and $p_0 = 0$ must be used within  the gauge-field propagator, such that the electron-electron interaction becomes exactly the Coulomb potential in (2+1)D \cite{Vozmediano1994, DasSarma2014}. Indeed, integrating out $A_\mu$ in Eq. \eqref{L_isotropic}, and taking the static limit, yields an effective action to the matter field \cite{Marino1993}. Therefore, the Lagrangian below represents the effective interaction between the electrons restricted in the plane in the static limit, namely,
\begin{equation}\label{L_eff}
\begin{split}
\mathcal{L}_{{\rm eff}} =& \, \, \bar{\psi_a}(i \gamma^{0}\partial_{0} +i v_{F}\gamma^{i} \partial_{i} - m)\psi_a \\
&+  \psi^{\dagger}_a (\bold x)\psi_a (\bold x) \frac{e^2}{4 \pi \varepsilon |\bold x - \bold y|} \psi^{\dagger}_a (\bold y)\psi_a (\bold y) \, .
\end{split}
\end{equation}
Furthermore, the gauge-field propagator in the Feynman gauge ($\xi = 1$) reads 
\begin{equation}\label{GaugeField_eff}
	\Delta_{00} (|\bold p|) = \frac{1}{2\varepsilon |\bold p|}\, .
\end{equation}
Note that the Coulomb interaction in Eq. \eqref{L_eff} is, essentially, the Fourier transform of Eq. \eqref{GaugeField_eff}.


\section{The renormalization of mass and Fermi velocity}\label{Sec_R_without_T}

In this section, we calculate the electron self-energy and derive the equations for both the renormalized Fermi velocity $v^R_F(p)$ and the renormalized mass $m^R(p)$ and compare our theoretical result for $m^R(p)$ with a few experimental data. 
Although the result for $v^R_F(p)$ is well known for $m=0$ \cite{Vozmediano1994}, the result for the renormalized mass has not been calculated within perturbation theory yet. 


According to the Feynman rules, the electron self-energy at one-loop order, is given by
\begin{equation}\label{Self-Energy}
\begin{split}
	\Sigma (\bold p) &= e^2 \mu^\epsilon \int \frac{d^d \bold k}{(2\pi)^{d}}  \frac{d k_0}{2\pi} \gamma^{0} S_{0F} ( \bar p - \bar k) \gamma^{0} \Delta_{00} (\bold k) \\
	&=\frac{e^2 \mu^\epsilon }{4\varepsilon}\int \frac{d^d \bold k}{(2\pi)^{d}}\frac{v_F \pmb{\gamma}.(\bold{p} - \bold{k}) - m}{|\bold k| \sqrt{v^2_F|\bold{p} - \bold{k}|^2 + m^2}} \, ,
\end{split}
\end{equation}
where we use the dimensional regularization, such that the $\epsilon$ is the regularization parameter, $\mu$ is an energy scale, and $d=2-\epsilon$ is the spatial dimension. This self-energy is shown in Fig.~\ref{Fig_SelfEnergy}. Note that for deriving the last line in Eq.~\eqref{Self-Energy}, we have used $\gamma^{0}\gamma_{0}=-1$ and $\gamma^0\gamma_i\gamma_0=\gamma_i$.

In order to solve the loop integral in Eq.~\eqref{Self-Energy}, we apply Feynman's parameterization and follow the standard procedure. Hence,
\begin{equation}
\begin{split}
\Sigma (\bold p) =& \,\, \alpha \mu^\epsilon \int^1_0 dx \frac{v_F\pmb{\gamma}.\bold{p}(1-x) - m}{x^{1/2} (1-x)^{1/2}} \\
&\times \int \frac{d^d k}{(2\pi)^d} \frac{1}{ \bold{k}^2 + \hat \Delta} ,
\end{split}
\end{equation}
where $\alpha  = e^2 / 4\pi \varepsilon v_F$ is the fine-structure constant and $\hat \Delta =  \bold{p}^2 x (1-x) + m^2 x/v^2_F$. From now on, we use $p$ to write the moment vector modulus, i.e., $p=|\bold p|$.

Next, we would like to compare our theoretical results with some experimental findings for two-dimensional materials, which have a typical lattice parameter $a$. It turns out that, in this case, one must consider an ultraviolet cutoff $\Lambda\propto 1/a$, where $a$ is related to the honeycomb lattice. Therefore, it is appropriate to make an association between the divergent terms in different regularization schemes at one-loop order \cite{Vozmediano1994}, which may be performed by doing
\begin{equation}
\frac{1}{\epsilon} \to \ln \left( \frac{\Lambda}{p} \right) .
\end{equation}
It is worth to mention that because we are describing electrons in a two-dimensional material, hence, the cutoff is finite and, therefore, the renormalized parameters shall be dependent on this parameter. Having in mind that $a \approx 1 \rm \AA $ for graphene, it follows that $\Lambda \approx 1 \rm e \rm V$, which establishes an upper-energy limit for the validity of the Dirac approximation for the quasi-particle. Finally, after solving the loop integral, we obtain the divergent term of the electron self-energy, given by
\begin{equation}\label{Sigma_Div}
\Sigma (\bold{p}) = \frac{\alpha }{4} \left[v_F \pmb{\gamma}.\bold{p} - 2m \right] \ln\left( \frac{\Lambda}{p} \right).
\end{equation}

Next, we consider the Schwinger-Dyson equation for the full electron propagator, given by
\begin{equation}\label{FullFermiProp}
  S^{-1}_F(p_0, \bold p) = S^{-1}_{0F}(p_0, \bold p) - \Sigma(\bold p)  \, .
\end{equation}
The quantum correction provided by $\Sigma(\bold p)$ modifies the parameters $v_F$ and $m$ in Eq.\eqref{FullFermiProp}, such that we obtain $ S^{-1}_F(p_0, \bold p) =  \gamma^0 p_0 + v^R_F \pmb \gamma . \bold p - m^R$, where $S^{-1}_{0F}(p_0, \bold p)$ and $\Sigma(\bold p)$ are given by Eq.~\eqref{ElectronPropagator} and \eqref{Sigma_Div}, respectively. After using this procedure, we find that
\begin{equation}
\begin{split}\label{Fermi_R}
\frac{v_F^R (\bold p)}{v_F} =&  1 + \frac{\alpha}{4} \ln \left( \frac{\Lambda}{p} \right)  \, , 
\end{split}
\end{equation}
which is in agreement with Ref.~\cite{Vozmediano1994}. This explains the renormalization of Dirac cones in Ref.~\cite{Elias2011}. On the other hand, the renormalized mass is given by
\begin{equation}\label{MassR}
\begin{split}
\frac{m^R(\bold p)}{m} &=  1 +  \frac{\alpha}{2} \ln \left( \frac{\Lambda}{p} \right) .
\end{split}
\end{equation}
This has been calculated in Ref.~\cite{Luis2020} in the light of the large-$N$ approximation.
\begin{figure}[h!]
	\centering
	\includegraphics[width=0.48\textwidth]{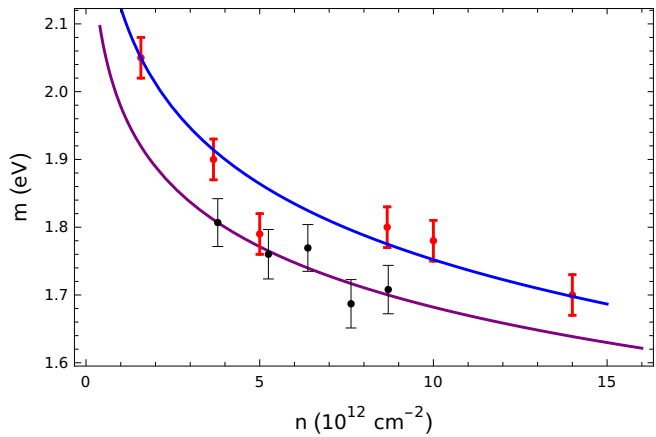}
	\caption{ The renormalization of the WSe$_2$ energy gap. The experimental data were extracted from Ref.~\cite{Fig4_WSe2}, measured at a temperature of 100 K, where two different substrates were used. The red dots correspond to the gap measurements on substrate $1$ and the black dots correspond to the measurements on substrate $2$. The blue and purple curves are obtained from Eq.~\eqref{MassR_n}. For the blue curve we use $\alpha = 0.31$, $m(n_0)=2.04 \, $ eV and $n_0 = 1.6\times 10^{-12} $ cm$^{-2}$  while for the purple curve we have $\alpha= 0.29$, $m(n_0)=1.80 \, $eV and $n_0 = 3.79\times 10^{-12} $ cm$^{-2}$. }\label{Fig_WSe2}
\end{figure}

Next, we compare the renormalized mass, described in Eq.~\eqref{MassR}, with the experimental measurements of the energy gap in WSe$_2$, measured in Ref.~\cite{Fig4_WSe2} and MoS$_2$, described in Ref.~\cite{Fig4_MoS2}. Similar to what has been done for the Fermi velocity, in order to perform the comparison between Eq.~\eqref{MassR} and the experimental results, we must replace the energy ratio $\Lambda/p$ to a ratio of carrier concentrations $n$. Therefore, $\Lambda/p \to (n_ 0/n)^{1/2}$ where $p$ is taken as the Fermi energy and $n_0$ is an arbitrary electronic density. This relation is a consequence of the fact that the Fermi energy of an ensemble of two-dimensional electrons is given by $p \sim n^{1/2}$. Furthermore, $\Lambda$ may be taken as an energy scale related to a fixed carrier concentration $n_0$ \cite{MarinoBook}. Hence,
\begin{equation}\label{MassR_n}
\begin{split}
\frac{m^R(n)}{m (n_0)} &=  1 +  \frac{\alpha}{4} \ln \left( \frac{n_0}{n} \right) .
\end{split}
\end{equation}

\begin{figure}[h!]
	\centering
	\includegraphics[width=0.48\textwidth]{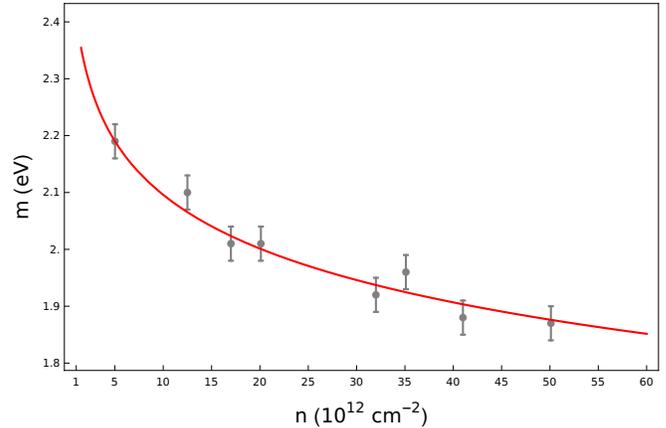}
	\caption{ The behavior of the  MoS$_2$ band gap normalization. Experimental data were extracted from \cite{Fig4_MoS2}, measured at a temperature of 295 K. We plot Eq.~\eqref{MassR_n} using $\alpha = 0.25$, $m(n_0)= 2.18 \, eV$ and $n_0 = 5.02 \times 10^{-12} cm^2$. }\label{Fig_MoS2}
\end{figure}

Eq.~(\ref{MassR_n}) shows that the renormalization of the energy band gap is measured by changing the carrier concentration. The renormalization measured in Ref.~\cite{Fig4_WSe2} was made by putting WSe$_2$ in two different substrates of boron nitride. These substrates have each one a different dielectric constant $\varepsilon$ and a different fine-structure constant ($\alpha \propto 1/\varepsilon$). Fig.~\ref{Fig_WSe2} shows the experimental dots with the erros bar extracted from Fig.~4 in Ref.~\cite{Fig4_WSe2}. The red dots refer to samples on substrate 1 with $d_{{\rm BN}} \approx\, 7.4 $ nm and black dots refer to samples on substrate 2 with $d_{{\rm BN}} \approx \, 4.5 $ nm. For the bare mass, we use $m(n_0)=2.04 \, $eV at $n_0 = 1.6\times 10^{-12} $cm$^{-2}$ and $m(n_0)=1.80 \, $eV at $n_0 = 3.79\times 10^{-12} $cm$^{-2}$  for red and black dots, respectively.
Thereafter, we use Eq.~\eqref{MassR_n} to find the best fit for the experimental points by using $\alpha$ as a free parameter, hence, in the blue curve we have $\alpha = 0.31$ and for the purple curve $\alpha= 0.29$. In Fig.~\ref{Fig_MoS2}, we repeat the same procedure for MoS$_2$ using the experimental data from  Fig. 4 in Ref.~\cite{Fig4_MoS2}. In this case, we have $m(n_0)= 2.18 \, $eV, $n_0 = 5.02 \times 10^{-12}$ cm$^{-2}$, and $\alpha = 0.25$. As we may conclude from these results, our renormalized parameters are in good agreement with these experimental data.

It is worth to mention that such agreement has already been discussed in Ref.~\cite{Luis2020}, where PQED is considered within the RPA to calculate the renormalized energy gap. In this case, it was obtained that $\alpha = 1.22$ and $\alpha = 0.97 $ for  WSe$_2$, placed in substrate 1 and 2, respectively, and $\alpha = 0.80$ for MoS$_2$. Because Ref.~\cite{Luis2020} have used a large-$N$ expansion (the same as the RPA), hence, an independent experimental measurement of $\alpha$ for these samples would establish what is the more accurate approach. This is yet to be done for the best of our knowledge. Regardless of this, the qualitative results are clearly consistent.

\section{The electron self-energy at finite temperature}\label{The_self-energy_at_T}

In this section, we calculate the effects of a thermal bath at temperature $T$. In order to do so, we consider the Matsubara formalism \cite{Matsubara, DasBookThermo}, hence, we rewrite the propagators in Eqs.~\eqref{ElectronPropagator} and \eqref{GaugeField}, using $p_0 \to \omega_l =(2l+1)\pi T$ in the fermion propagator, and $k_0 \to \omega_n = 2n\pi T$ in the gauge-field propagator, where $(l,n)$ are integers. Therefore,
the loop integrals have its time-component integration converted into a sum, for example,
\begin{equation}
\int_{-\infty}^{+\infty} dk_0 I(k_0,\textbf{k}) \to 2\pi T \sum^\infty_{n=-\infty} I(\omega_n,\textbf{k}),
\end{equation}
where $I(k_0,\textbf{k})$ is an arbitrary integrand. This completes our Feynman rules for including finite temperature effects.

Having these properties in mind and considering the static limit, the electron self-energy described in Fig.~\ref{Fig_SelfEnergy} reads
\begin{equation}
\begin{split}
\Sigma_l (T, \bold{p}) =& \,\, T e^2 \int \frac{d^2 k}{(2\pi)^2} \sum^\infty_{n= - \infty} \gamma^0  \\ 
&\times S_F (\omega_l - \omega_n, \bold p -\bold k) \gamma^0 \Delta_{00} (\bold k). 
\end{split}
\end{equation}
Furthermore, for the sake of simplicity, we calculate the zero mode $l=0$ of the self-energy, which is the most relevant contribution of this amplitude \cite{Leandro}. Hence,
\begin{equation}
\begin{split}
\Sigma (T, \bold{p}) =& \,\, - \frac{e^2}{2\varepsilon } T \int \frac{d^2 k}{(2\pi)^2} \frac{1}{|\bold{k}|} \left\lbrace \gamma^0\pi T S_1 \right. \\
&\left. - \left[v_F \pmb \gamma . (\bold p - \bold k) - m\right] S_2 \right\rbrace,
\end{split}
\end{equation}
where
\begin{equation}\label{Sum_1}
\begin{split}
S_1 &= \sum^\infty_{n=-\infty} \frac{1-2n}{ \left[ (-2n + 1)^2\pi^2T^2 + E(\bold p,\bold k, v_F, m)^2  \right]} \\
&= 0,
\end{split}
\end{equation}
and
\begin{equation}\label{Sum_2}
\begin{split}
S_2 &= \sum^\infty_{n=-\infty} \frac{1}{ \left[ (-2n + 1)^2\pi^2T^2 + E(\bold p,\bold k, v_F, m)^2  \right]} \\  
&= \frac{\tanh \left( \frac{E(\bold p,\bold k, v_F, m)}{2T}\right)}{2TE(\bold p,\bold k, v_F, m)} ,
\end{split}
\end{equation}
where $E^2(\bold p,\bold k, v_F, m) = v^2_F|\bold{p} - \bold{k}|^2 + m^2$ (see Appendix~\ref{App_Summation} for a detailed calculation of $S_1$ and $S_2$). Therefore, we obtain
\begin{equation}\label{SigmaTanH}
\begin{split}
\Sigma (T, \bold{p}) =& \,\, \frac{e^2}{4 \varepsilon} \int \frac{d^2 k}{(2\pi)^2} \frac{v_F \pmb \gamma . (\bold p - \bold k) - m}{|\bold{k}| E(\bold p,\bold k, v_F, m)}  \\
&\times \tanh \left( \frac{ E(\bold p,\bold k, v_F, m) }{2T}\right).
\end{split}
\end{equation}

It is convenient to use the identity $\tanh(x) = 1 - 2\,n_F (2x)$ in Eq.~\eqref{SigmaTanH}, where $x = E(\bold p,\bold k, v_F, m) / 2T $ and
\begin{equation}\label{Nf}
n_F(T,\bold p, \bold k, v_F, m)=\frac{1}{\exp\left(\frac{E(\bold p,\bold k, v_F, m)}{T}\right) + 1 },
\end{equation}
which is the Fermi-Dirac distribution. This is useful for separating the zero-temperature contribution from the finite-temperature term, i.e,
\begin{equation}\label{Sigma_T0_e_T}
\Sigma (T, \bold{p}) = \Sigma (0, \bold{p}) + \hat \Sigma (T, \bold{p}),
\end{equation}
where
\begin{equation}\label{Sigma_T0}
 \Sigma (0, \bold{p}) =  \alpha \pi v_F  \int \frac{d^2 k}{(2\pi)^2} \frac{v_F \pmb{\gamma.}(\bold{p} - \bold{k}) - m}{|\bold{k}| E(\bold p,\bold k, v_F, m)}
\end{equation}
is the zero-temperature term, calculated in Ref.~\cite{Vozmediano1994}. This is the same expression represented by Eq.~\eqref{Self-Energy}. On the other hand,
\begin{equation}\label{Sigma_T}
\begin{split}
\hat \Sigma (T, \bold{p}) =& - 2\alpha \pi v_F  \int \frac{d^2 k}{(2\pi)^2} \frac{v_F \pmb{\gamma}.(\bold{p} - \bold{k}) - m}{|\bold{k}| E(\bold p,\bold k, v_F, m)} \\
&\times n_F (T,\bold p,\bold k, v_F, m) \, ,
\end{split}
\end{equation}
is the finite-temperature term. Accordingly to Eq.~\eqref{Nf}, the Fermi-Dirac distribution vanishes when $T \to 0$. However, $n_F\rightarrow 1/2$ when $T \to \infty$. In this sense, it follows that $\Sigma (T\rightarrow\infty, \bold{p})\rightarrow 0$, which implies that the parameters of the model are not renormalized by quantum corrections and, therefore, they remain with the same value given by the noninteracting theory whenever $T \to \infty$. Next, let us calculate the temperature-dependent term of the electron self-energy.

\subsection{The Angular Integral}\label{TheAngularIntegral}
The first step is to solve the angular integral given in Eq.~\eqref{Sigma_T} by using $\bold{k} \rightarrow \bold{k}+\bold{p}$. Thereafter, we assume polar coordinates, where $d^2k=kdk d\theta$ with $0\leq \theta \leq 2\pi$ and $0\leq k \leq \Lambda$. Our variable change from $(k_x, k_y)$ to $(k, \theta)$ implies that: $k_x = k\cos\theta$ and $k_y = k\sin\theta $. Next, we take advantage of the polar symmetry of the electron self-energy with respect to the external momentum $\bold{p}=(p_x,p_y)$. This allows us to consider, without loss of generality, that $ \bold{p}=(p,0)$ which simplifies some of our calculations. Following these assumptions, we find
\begin{equation}\label{Sigma_Int_k}
\begin{split}
\hat \Sigma (T, p) =& \,\, \frac{\alpha v_F}{2\pi}  \int_0^{\Lambda} dk \frac{n_F (T, k, v_F, m)}{E (k, v_F, m)}  \\ 
&\times \left[\gamma_x k^2 I_4(k,p) + \gamma_y k^2 I_5(k,p) \right. \\
&\left. + k I_6(k,p) m \right].
\end{split}
\end{equation}
The angular integrals are given by
\begin{equation}
\begin{split}
\frac{I_4(k,p)}{v_F} =& \,\, \int^{2\pi}_0 d\theta \frac{\cos\theta}{ \sqrt{k^2 + p^2 + 2pk\cos\theta} } \\
=& \, \, \frac{2(k+p)}{k \, p} \left[ \mathbb{E} \left( \frac{4\, k \,p}{(k+p)^2}\right) \right.\\ &\left. - \frac{(k^2+p^2)}{(k+p)^2} \mathbb{K} \left( \frac{4\, k \,p}{(k+p)^2}\right)  \right] \, , \\
\frac{I_5(k,p)}{v_F} =& \,\,  \int^{2\pi}_0 d\theta \frac{\sin\theta}{ \sqrt{k^2 + p^2 + 2pk\cos\theta} }\\
=& \,\, 0  \, ,\\
\frac{I_6(k,p)}{m } =&  \int^{2\pi}_0 d\theta \frac{1}{ \sqrt{k^2 + p^2 + 2pk\cos\theta} } \\
=& \,\, \frac{4}{(k+p)} \mathbb{K} \left( \frac{4\, k \,p}{(k+p)^2}\right),
\end{split}
\end{equation}
where the functions $\mathbb{E}$ and $\mathbb{K}$ are the complete elliptic integral of the second and first  kind, respectively \cite{Grads}. Hence, after we recover the $p_y$-component due to the rotational symmetry in the $p_x$-$p_y$ plane, the electron self-energy in Eq.~\eqref{Sigma_Int_k} reads
\begin{equation}\label{SigmaHat_Func_f_g}
\begin{split}
\hat \Sigma (T, \bold{p}) =& \,\,\frac{\alpha v_F}{ \pi} \left\lbrace v_F \pmb{\gamma}.\textbf{p} \int_0^{\Lambda} dk \, \frac{f(p,k) n_F(T,k, v_F, m)}{E (k, v_F, m)} \right. \\   
&\left.+ 2 m  \int_0^{\Lambda} dk \, \frac{g(p,k) \,  n_F(T,k, v_F, m)}{{E (k, v_F, m)}} \right\rbrace,
\end{split}
\end{equation}
where
\begin{equation}\label{f}
\begin{split}
f(p,k) =& \,\, \frac{k \,(k+p)}{p^2} \left[ \mathbb{E} \left( \frac{4\, k \,p}{(k+p)^2}\right) \right. \\
&\left. - \frac{k^2+p^2}{(k+p)^2} \mathbb{K}\left( \frac{4\, k \,p}{(k+p)^2}\right)  \right],  
\end{split}
\end{equation}
and
\begin{equation}\label{g}
\begin{split}
g(p,k) =& \frac{k}{(k+p)} \mathbb{K} \left( \frac{4\, k \,p}{(k+p)^2}\right) .
\end{split}
\end{equation}
Note that, from Eq.~(\ref{SigmaHat_Func_f_g}), we may schematically write the electron self-energy as
\begin{equation}\label{Sigma_F_G}
\hat \Sigma (T, \bold{p})=\alpha v_F \pmb \gamma . \bold p \, F(T, p) - \alpha m  \, G (T, p).
\end{equation}
This decomposition shall be useful for calculating both the renormalized Fermi velocity and the renormalized electron mass.

\subsection{The Low-temperature Regime}

In this case, the Fermi-Dirac distribution $n_F$ is written as
\begin{equation}
\begin{split}
n_F =&\,\, \exp\left(\frac{- E (k, v_F, m)}{T}\right)  \\
&\times \left[ \frac{1}{1+\exp\left(\frac{-E (k, v_F, m)}{T}\right) } \right],
\end{split}
\end{equation}
which, in the low-temperature regime, is conviniently written as
\begin{equation}
n_F = \sum^\infty_{j=0} (-1)^j \exp\left[ \frac{-(1+j) E (k, v_F, m)}{T}\right].
\end{equation}

In order to solve the integral over $k$ in Eq.~(\ref{SigmaHat_Func_f_g}), we consider
\begin{equation}\label{Split_Integral}
\int^\Lambda_0 dk  I(k,p) = \int^p_0 dk I(k,p) + \int^\Lambda_p  dk  I(k,p) \, ,
\end{equation} 
where $I(k,p)$ is an arbitrary function. Furthermore, in the region $ k\in [0,p]$, we consider $p\gg k$ and, in the region $k\in [p, \Lambda]$, we use $k\gg p$. Hence, the functions $f(p,k)$ and $g(p,k)$, given by Eqs.~\eqref{f} and \eqref{g}, may be written as
\begin{equation}\label{fSerie2}
f(k,p) = \sum^{\infty}_{n=0} \mathcal B_{2n+3} \left( \frac{k^{2n+3}}{p^{2n+3}}  \Theta (p-k) + \frac{p^{2n}}{k^{2n}}  \Theta (k-p)  \right)
\end{equation}
and
\begin{equation}\label{gSerie2}
g(k,p) = \sum^{\infty}_{q=0} \mathcal C_{2q+1} \left( \frac{k^{2q+1}}{p^{2q+1}} \Theta (p-k) + \frac{p^{2q}}{k^{2q}}  \Theta (k-p)  \right) \, ,
\end{equation} 
where $(\mathcal B_{2n+3},\mathcal C_{2q+1})$ are known constants and $\Theta (k)$ is the Heaviside function. For more details regarding these expansions, please see Appendix~\ref{App_SerieExp}. 

Finally, using these assumptions, the functions $F(T,p)$ and $G(T,p)$, see Eq.~\eqref{Sigma_F_G}, are given by
\begin{equation}\label{FTP}
\begin{split} 
F (T, p) =& \frac{ v_F}{ \pi}  \sum^{\infty}_{j=0}\sum^{\infty}_{n=0}  (-1)^j \mathcal B_{2n+3} \times \\
&\times \left\lbrace  \frac{1 }{p^{2n + 3}} I_{r_1} + p^{2n} I_{z_1} \right\rbrace
\end{split}
\end{equation}
and
\begin{equation} \label{GTP}
\begin{split}
G (T, p) =& - 2 \frac{ v_F}{ \pi} \sum^{\infty}_{j=0} \sum^{\infty}_{q=0} (-1)^j \mathcal C_{2q+1} \times \\
&\times \left\lbrace  \frac{1}{p^{2q+1}} I_{r_2} + p^{2q} I_{z_2} \right\rbrace ,
\end{split}
\end{equation}
where we have defined four definite integrals, given by
\begin{equation}
\begin{split}
I_r = \int_0^{p} dk \, \frac{k^r \,  \exp\left( \frac{-E (k, v_F, m)}{\mathcal{T}_j}\right)}{E (k, v_F, m)}
\end{split}
\end{equation}
and
\begin{equation}
\begin{split}
I_z = \int_p^{\Lambda} dk \, \frac{ \exp\left( \frac{-E (k, v_F, m)}{\mathcal{T}_j}\right)}{k^z E (k, v_F, m)} \, ,
\end{split}
\end{equation}
where $r$ and $z$ may assume two values, namely, $( r_1 = 2n+3, \, r_2 = 2q+1)$ for $r$ and  $( z_1 = 2n , \, z_2 = 2q)$ for $z$. On the other hand, $\mathcal{T}_j = T/(1+j)$ encodes the temperature. Note that both $I_r$ and $I_z$ quickly go to zero as we increase $m$. Therefore, in order to calculate an analytical solution, we may take only the leading term in $I_r$ and $I_z$, which is obtained by using $m\rightarrow 0$ within the integrand. Obviously, this is only a practical approximation for the analytical result and we shall relax this condition in the numerical tests. Hence, we find
\begin{equation}
\begin{split}
I_{r} &= \int_0^{p} dk \, k^r \,  \left(\frac{e^{-v_F k/\mathcal{T}_j}}{v_F k}\right)  \\
&= \frac{1}{v_F} \left(\frac{\mathcal{T}_j}{v_F}\right)^r \left[ \Gamma(r) -\Gamma\left( r, \frac{v_F p}{\mathcal{T}_j} \right) \right]
\end{split}
\end{equation}
and
\begin{equation}
\begin{split}
I_{z} &= \int_p^{\Lambda} dk \, \frac{ 1 }{k^z}  \left(\frac{e^{-v_F k/\mathcal{T}_j}}{v_F k} \right)\\
&= \frac{1}{v_F} \left(\frac{\mathcal{T}_j}{v_F}\right)^{-z} \left[ \Gamma\left( z, \frac{v_F p}{\mathcal{T}_j} \right) -\Gamma\left( z, \frac{v_F \Lambda}{\mathcal{T}_j} \right) \right], 
\end{split}
\end{equation}
where $\Gamma(N,\chi)$ is the incomplete gamma function, which admits the following expansion \cite{Grads}
\begin{equation}\label{Gammaexp}
\Gamma(N, \chi) =  (N - 1)  ! \, e^{-\chi} \sum^{N - 1}_{s = 0} \frac{\chi^s}{s !}.
\end{equation}

Because we are considering a low-temperature expansion, we may consider only the lowest-order term in Eq.~(\ref{Gammaexp}). Using this approximation in Eq.~(\ref{FTP}) and Eq.~(\ref{GTP}), we find, after some algebra, our final expressions for both $F(T,p)$ and $G(T,p)$, namely,
\begin{equation} \label{FTP2}
\begin{split}
F (T, p) =&\,\, \frac{1}{ \pi} \sum^{\infty}_{j=0}\sum^{\infty}_{n=0}  (-1)^j \mathcal B_{2n+3}  \\
& \times \left( \frac{\mathcal{T}_j }{ v_F p} \right)^{2n + 3} \left\lbrace   \Gamma(2n+3) \right. \\
&\left.  - \left( \frac{\mathcal{T}_j }{ v_F p} \right)^{-4n - 3} \left( \frac{\mathcal{T}_j }{ v_F \Lambda} \right)^{2n +1} e^{- v_F \Lambda / \mathcal{T}_j} \right\rbrace
\end{split}
\end{equation}
and
\begin{equation}\label{GTP2}
\begin{split}
G (T, p) =& \,\, - \frac{2}{ \pi} \sum^{\infty}_{j=0} \sum^{\infty}_{q=0} (-1)^j \mathcal C_{2q+1} \\
& \times \left( \frac{\mathcal{T}_j }{ v_F p} \right)^{2q + 1} \left\lbrace   \Gamma(2q+1)  \right. \\
&\left. - \left( \frac{\mathcal{T}_j }{ v_F p} \right)^{-4q-1} \left( \frac{\mathcal{T}_j }{ v_F \Lambda} \right)^{2q+1} e^{- v_F\Lambda / \mathcal{T}_j} \right\rbrace \, ,
\end{split}
\end{equation}
which may be applied in Eq.~(\ref{Sigma_F_G}).

From Eq.~(\ref{FTP2}) and Eq.~(\ref{GTP2}), we may calculate the most relevant contribution for $n$ and $q$, then we sum over $j$. Furthermore, using the known values of the constants $\mathcal B_3=-\mathcal C_1=-\pi/2$ (see Appendix~\ref{App_SerieExp}), we may conclude that
\begin{equation}\label{F_Result}
\begin{split}
F (T,p)=& - \frac{3\, \zeta(3)}{4}  \left( \frac{T }{ v_F p} \right)^{3} + O \left[ \left( \frac{T }{ v_F p} \right)^{5} \right]
\end{split}
\end{equation}
and
\begin{equation}\label{G_Result}
\begin{split}
G (T,p)=& - \frac{T \ln 2 }{ v_F p}+ O \left[ \left( \frac{T }{ v_F p} \right)^{3} \right],
\end{split}
\end{equation}
where $\zeta(x)$ is the zeta function with $\zeta(3) \simeq 1.202$.

Let us discuss the validity of our analytical approximation. For a graphene-like system, we must consider $m\rightarrow 0$ in Eq.~\eqref{SigmaTanH}, whose main effect is that only the $F$-term in Eq.~(\ref{F_Result}) does not vanish, yielding the $v_F$ renormalization, as it has been shown in Ref.~\cite{Vozmediano1994} at zero temperature. In this case, the physical cutoff is in order of $\Lambda \approx 1$eV, which also sets an upper-energy limit for the validity of the Dirac-like description, hence, we may expect that $v_F p\ll \Lambda\approx 1$eV. On the other hand, we have assumed that $k_B T\ll v_F p$ (recovering the physical value of the Boltzmann constant $k_B\approx 10^{-4}$eV/K). Therefore, assuming that the kinetic energy of the electrons in graphene is not larger than one-tenth of $\Lambda$, we conclude that our approximation should work well for temperatures much less than $10^3$K, such as the room temperature $~10^2$K. Indeed, it has been shown that even the zero-temperature limit already provides a reasonable comparison for the renormalization of the Fermi velocity. Regardless of this estimative, note that in Eq.~(\ref{SigmaHat_Func_f_g}) we actually have an analytical result for any temperature and bare mass.

\section{The renormalization at finite temperature}\label{The_Renormalization_at_T}

In this section, we show how to obtain the renormalized parameters using the electron self-energy. From the Schwinger-Dyson equation for the full electron propagator, we find
\begin{equation}\label{SDE_2}
S^{-1}_{0F}(\omega, \bold p) - \Sigma(\omega,\bold p) =  \gamma^0 \omega + v^R_F \pmb \gamma . \bold p - m^R  \, ,
\end{equation}
where $S^{-1}_{0F}(\omega, \bold p)$ is the bare fermion propagator that contains Matsubara frequencies and $\Sigma(\omega,\bold p)$ is described in Eq.~\eqref{Sigma_T0_e_T}. Within the low-temperature approximation, the temperature-dependent term of the electron self-energy is written in Eq.~\eqref{Sigma_F_G}, whose coefficients are given by Eq.~\eqref{F_Result} and Eq.~\eqref{G_Result}. In addition, the zero-temperature term  is calculated in Sec.~\ref{Sec_R_without_T}, see Eq.~\eqref{Sigma_Div}. Under these circumstances, after using these results in Eq.~(\ref{SDE_2}), we may identify the renormalized Fermi velocity as
\begin{equation}\label{FermiV_SmallT}
\begin{split}
\frac{v_F^R (T, \bold p)}{v_F} =& \,\, 1 + \alpha \left\lbrace \frac{1}{4} \ln \left( \frac{\Lambda}{p} \right) - \frac{3\, \zeta(3)}{4} \left(\frac{T}{v_F p}\right)^3 \right. \\
&+\left.  \mathcal{O} \left[ \left(\frac{T}{v_F p} \right)^5 \right] \right\rbrace  \,.
\end{split}
\end{equation}
On the other hand, the renormalized mass reads
\begin{equation}\label{Mass_SmallT}
\begin{split}
\frac{m^R(T, \bold p)}{m} =& \,\, 1 + \alpha \left\lbrace \frac{1}{2} \ln \left( \frac{\Lambda}{p} \right) - \left(\frac{T \ln 2}{v_F p}\right) \right. \\
&+\left. \mathcal{O} \left[ \left(\frac{T}{v_F p} \right)^3 \right]   \right\rbrace \, .
\end{split}
\end{equation}
From these results, we conclude that the temperature acts as an inhibition of the renormalization, however, this effect is weaker in the Fermi velocity, as it occurs at order of $T^3$ in comparison to the linear term in the renormalization of the mass. Interesting, an inhibition behavior in the Fermi velocity renormalization has been investigated in Ref.~\cite{Danilo2017} due to the presence of a conducting plate.

\begin{figure}[h!]
	\centering
	\includegraphics[width=0.48\textwidth]{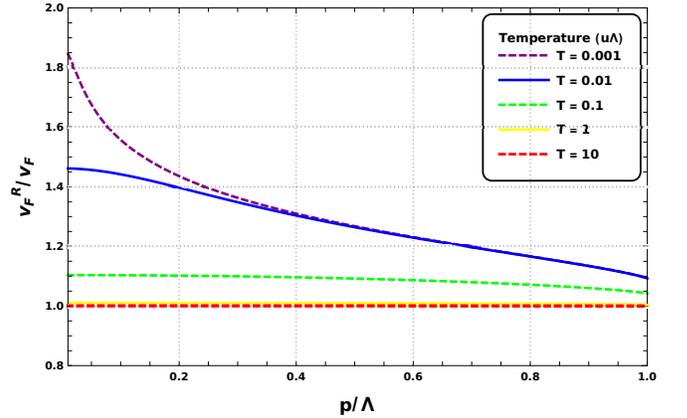}
	\caption{The ratio between the renormalized Fermi velocity to the bare Fermi velocity as a function of external momentum $p$. The curves are produced from Eq.~\eqref{FermiV_NI} considering $m=0 \,u \Lambda$, $v_F=3.7/300$, $\alpha = 0.7$ and $\Lambda = 10\,u \Lambda$.}\label{Fig_vr_p-T}
\end{figure}

\begin{figure}[h!]
	\centering
	\includegraphics[width=0.48\textwidth]{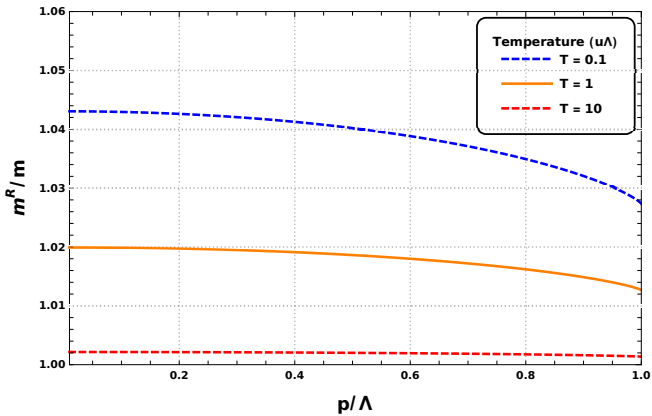}
	\caption{The ratio of the $m^R$ and $m$ as a function of external momentum $ p$. We plot Eq.~\eqref{Mass_NI} with $m=1 \,u \Lambda$, $v_F=3.7/300$, $\alpha = 0.7$ and $\Lambda = 10 \,u \Lambda$.}\label{Fig_mr_p-T}
\end{figure}

We can improve these previous results regarding the influence of the thermal bath on the renormalized parameters. For this, we calculate the integral equations for these parameters from Eq.~\eqref{SDE_2} in which the temperature-dependent term of the electron self-energy, in Eq.~\eqref{Sigma_T0_e_T}, is given by Eq.~\eqref{SigmaHat_Func_f_g} while the temperature-independent term is derived from Eq.~\eqref{Sigma_T0}, written in angular variables.
Therefore, the renormalized equations for both the Fermi velocity and mass are
\begin{equation}\label{FermiV_NI}
\begin{split}
\frac{v^R_F(T, \bold p)}{v_F} =& \,\, 1 - \frac{\alpha v_F}{2\pi}\int^\Lambda_0 dk \frac{f(p,k)}{E(k,v_F,m)} \\
&+ \frac{\alpha v_F}{\pi}\int^\Lambda_0 dk \frac{f(p,k) n_F(T,k,v_F,m)}{E(k,v_F,m)} 
\end{split} 
\end{equation}
and
\begin{equation}\label{Mass_NI}
\begin{split}
\frac{m^R(T, \bold p)}{m} =& \,\, 1 + \frac{\alpha v_F}{\pi}  \int^\Lambda_0 dk \frac{g(p,k)}{E(k,v_F,m)} \\
&-\frac{2\alpha v_F}{\pi}  \int^\Lambda_0 dk \frac{g(p,k) n_F(T,k,v_F,m)}{E(k,v_F,m)} \, .
\end{split} 
\end{equation}
Obviously, we may use numerical integration for solving Eq.~\eqref{FermiV_NI} and Eq.~\eqref{Mass_NI} for $v^R_F$ and $m^R$, respectively.

Fig.~\ref{Fig_vr_p-T} and Fig.~\ref{Fig_mr_p-T} are generated from Eqs.~\eqref{FermiV_NI} and \eqref{Mass_NI}, respectively. They show the behavior of the renormalized parameters at different temperatures. We chose the values: $v_F=3.7/300$; $\alpha = 0.7$; and  $\Lambda = 10$ u$\Lambda$ (units of $\Lambda$).
Fig.~\ref{Fig_vr_p-T} shows the ratio between the renormalized Fermi velocity and the bare velocity as a function of $p/\Lambda$. We  also consider $m = 0$, which applies for a graphene-like system. 
As expected, when the external momentum is close to the cutoff, the effect of the renormalization  decreases, moreover, when the temperature increases the renormalization also decreases.
For higher temperatures, there is no renormalization, as we can conclude when $T>1$ u$\Lambda$ the ratio between the velocities is quite close to one. In general grounds, at very high-temperatures, we obtain $n_F(T,k,v_F,m) \to 1/2$ and the temperature-dependent term cancels the temperature-independent term of the electron self-energy. The Coulomb interaction is, therefore, irrelevant in such system.

Fig.~\ref{Fig_mr_p-T}~ shows the ratio between the renormalized mass and the bare mass as a function of the momentum $p$. We use the $m=1$ u$\Lambda$. As in the previous case, the normalized mass also decreases as we increase the temperature until to reach its bare value.  

An interesting self-consistency test we can do is to compare the integral equations with our low-temperature analytical results.
In this case, it is enough only to consider the temperature-dependent term divided by their bare parameters. Therefore, we can write the renormalized Fermi velocity in Eq.~\eqref{FermiV_SmallT} and Eq.~\eqref{FermiV_NI} as
\begin{equation}\label{Delta_vA}
\frac{\Delta v_F^{R} (T, \bold p)}{v_F} = - \alpha \frac{3\, \zeta(3)}{4} \left(\frac{T}{v_F p}\right)^3 
\end{equation}
and
\begin{equation}\label{Delta_vN}
\frac{\Delta v^{R}_F(T, \bold p)}{v_F} = \frac{\alpha v_F}{\pi}\int^\Lambda_0 dk \frac{f(p,k) n_F(T,k,v_F,m)}{E(k,v_F,m)} \, .
\end{equation}
Similarly, Eq.~\eqref{Mass_SmallT} and Eq.~\eqref{Mass_NI} are given by
\begin{equation}\label{Delta_mA}
\frac{\Delta m^{R}(T, \bold p)}{m} = -\alpha  \left(\frac{T \ln 2}{v_F p}\right) 
\end{equation}
and
\begin{equation}\label{Delta_mN}
\frac{\Delta m^{R}(T, \bold p)}{m} = -\frac{2\alpha v_F}{\pi}  \int^\Lambda_0 dk \frac{g(p,k) n_F(T,k,v_F,m)}{E(k,v_F,m)}  \, .
\end{equation}

\begin{figure}[h!]
	\centering
	\includegraphics[width=0.48\textwidth]{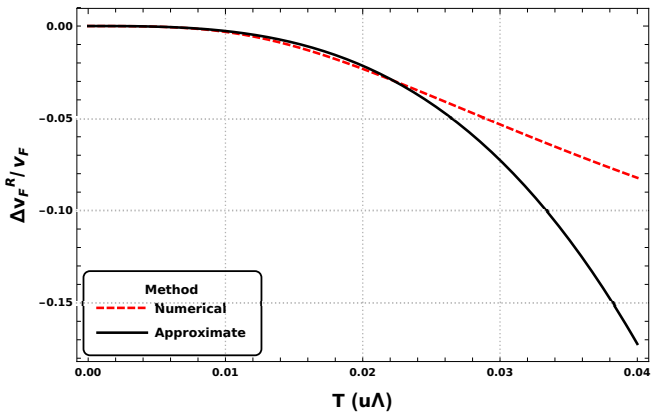}
	\caption{ Contribution of the temperature-dependent term to the renormalized Fermi velocity divided by the bare Fermi velocity. We plot Eqs.~\eqref{Delta_vA} and \eqref{Delta_vN} considering $m=0 \,u \Lambda$, $v_F=3.7/300$, $\alpha = 0.7$, $p=5 \,u \Lambda$, and $\Lambda = 10 \,u \Lambda$. The black line is the approximative result and the red dashed line is the numerical result.}\label{Fig_Comparison_v}
\end{figure}

\begin{figure}[h!]
	\centering
	\includegraphics[width=0.48\textwidth]{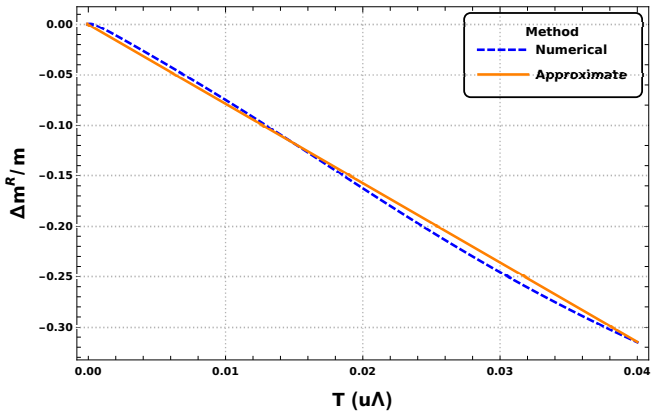}
	\caption{ The ratio between the temperature-dependent term of the renormalized mass and the bare mass. We consider $m=0.001 \,u \Lambda$, $v_F=3.7/300$, $\alpha = 0.7$, $p=5 \,u \Lambda$ and $\Lambda = 10 \,u \Lambda$. The orange line is obtained from Eq.~\eqref{Delta_mA} and the blue dashed line is obtained from Eq.~\eqref{Delta_mN}.}\label{Fig_Comparison_m}
\end{figure}

Fig.~\ref{Fig_Comparison_v} and Fig.~\ref{Fig_Comparison_m} show the comparison between approximate and numerical results. For both figures, we choose the values $v_F=3.7/300$, $\alpha = 0.7$, $p=5$ u$\Lambda$, and $\Lambda = 10$ u$\Lambda$.
Fig.~\ref{Fig_Comparison_v} shows the comparison between the Fermi velocities, in which, the analytical result in Eq.~\eqref{Delta_vA} is shown by the black line while the numerical result in Eq.~\eqref{Delta_vN} is shown by red dashed line. In this figure, we take $m=0$. From this plot, we can conclude that up to $T \simeq 0.024$ u$\Lambda$, our analytical result is in good agreement with the integral equation. Close to this point, the temperature-dependent term is around $3.5\%$ of the bare Fermi velocity.

Fig.~\ref{Fig_Comparison_m} also shows a good agreement between the approximated results (orange line), given by Eq.~\eqref{Delta_mA}, and the numerical results (blue dashed line), given by Eq.~\eqref{Delta_mN}. 
For Eq.~\eqref{Delta_mA}, which holds in the low-temperature regime, we choose $m=0.001$ u$\Lambda$. 
In these conditions, we can observe that within the range of temperature from $0$ to $0.04$ u$\Lambda$, the approximated solution coincides with the numerical results. Note that at $T=0.04$ u$\Lambda$, the temperature-dependent term is close to  $30\%$ of the bare mass, showing a relevant contribution. 
However, for $m>0.001$ u$\Lambda$ and $T\gg 0.04$ u$\Lambda$, our analytical approximations disagree with our numerical results, as expected.

\section{Discussions}\label{Discussions}

We have calculated both the Fermi velocity and the mass renormalization in PQED at finite temperature. These, for $T=0$, are consistent with the results in Ref.~\cite{Luis2020}, which have been calculated in the RPA approach and have been shown to be in agreement with the energy gap measurements in WSe$_2$ and MoS$_2$. Nevertheless, a critical difference is observed in the numerical values of the fine-structure constant for each monolayer. Therefore, we believe that an independent estimative for $\alpha$ would answer which is the more accurate approach. For example, one could use the excitonic spectrum in these two-dimensional materials \cite{Excitons2018} in order to obtain $\alpha$. Furthermore, a deeper investigation of the substrate role also seems to be relevant for providing a correction to our results. Here, we assume the simplest effect where the substrate only yields a screened Coulomb interaction through an effective dieletric constant.

In the low-temperature regime, we conclude that $v^R_F$ depends on $T^3$ and $m^R$ is linearly dependent on $T$. 
It is worth to mention that the effects of phonons on the optical band gap, for WSe$_2$, also shows the linear-temperature behavior accordingly to the experimental measurements in Ref.~\cite{Liu}. Here, however, we describe the contribution of the electron-electron interaction to the renormalization of the energy band gap, which has been experimentally measured by changing the electronic density. We also find integral equations for $v^R_F$ and $m^R$ for any temperature, within the static approximation, whose numerical solutions confirm our analytical results. Finally, it is expected that our theoretical results improve our understanding about the renormalized parameters in these 2D materials. Furthermore, it provides an interesting connection between a quantum-electrodynamical theory and a two-dimensional material.

\section*{Acknowledgement}
N. B. is partially supported by Coordenação de Aperfeiçoamento de Pessoal de Nível Superior Brasil (CAPES), finance code 001.

\appendix
\numberwithin{equation}{section}
\section{The Matsubara Sum over $n$}\label{App_Summation} 
\subsection{The first sum over $n$}
In order to solve the sum in Eq.~\eqref{Sum_1}, we write it as
\begin{equation}
\begin{split}
S_1 &= K^2 \sum^\infty_{n=-\infty} \frac{(1-2n)}{(1-2n)^2 + B^2},
\end{split}
\end{equation}
where
\begin{equation}\label{AKB}
\begin{split}
K &= \frac{1}{\pi T}; \\
B &= \frac{E}{\pi T};\\
E^2 &= v_F^2 |\bold{p} - \bold{k}|^2 + m^2.
\end{split}
\end{equation}
Thereafter, we split this sum in two parts: one referring to the sum over negative values of $n$ and the other over zero and the postitive values. Next, we make a variable change in the first part, given by $n \to - (n - 1)$, to obtain
\begin{equation}
\begin{split}
S_1 &= K^2 \sum^\infty_{n= 1} \left[ \frac{(1 - 2n)}{(1-2n)^2 + B^2} - \frac{(1-2n)}{(2n-1)^2 + B^2} \right]\\
&= 0,
\end{split}
\end{equation}
which clearly vanishes.

\subsection{The second sum over $n$}

We write the sum in Eq.~\eqref{Sum_2} as
\begin{equation}\label{S2}
\begin{split}
S_2 &= K^2 \sum^\infty_{n=-\infty} \frac{1}{(1-2n)^2 + B^2},
\end{split}
\end{equation}
where $B$ and $K$ are given by \eqref{AKB}. In order to solve this sum, we will do the same procedure as in previous section: divide the sum in two parts, one due to the negative values of $n$ plus the zero and positive values. Thereafter, we make a variable change, namely,  $- (n - 1)$, yielding 
\begin{equation}
\begin{split}
S_2 &=  2 K^2 \sum^\infty_{n= 1} \frac{1}{(2n - 1)^2 + B^2}.
\end{split}
\end{equation}
Then, using an identity \cite{Grads}
\begin{equation}
\tanh\left( \frac{\pi \lambda}{2} \right) = \frac{4 \lambda}{\pi}  \sum^\infty_{n=1} \frac{1}{(2n-1)^2 + \lambda^2},
\end{equation}
we may conclude that
\begin{equation}
S_2 = 2 K^2 \frac{\pi}{4 B} \tanh\left( \frac{\pi B}{2} \right).
\end{equation}
Using our constants, we find
\begin{equation}
S_2 = \frac{\tanh\left( \frac{E}{2T} \right)}{2 T E},
\end{equation}
which is the relevant result for calculating the electron-self energy.

\section{Some useful expansions}\label{App_SerieExp}

It is easy to show that the functions in Eq.~\eqref{f} and Eq.~\eqref{g} can be written in terms of a power series, given by
\begin{equation}\label{fSerie}
f(k,p) = 
\left\lbrace
\begin{split} 
 &-\frac{\pi k^3 }{2 p^3} - \frac{3\pi k^5 }{16 p^5} + \mathcal{O}\left( \frac{k^7}{p^7} \right) \, , \quad  p \gg k \\ 
 &- \frac{\pi }{2 } - \frac{3 \pi p^2}{16 k^2} + \mathcal{O}\left( \frac{p^4}{k^4} \right) \, , \quad k \gg p 
\end{split}
\right.
\end{equation}
and
\begin{equation}\label{gSerie}
g(k,p) = 
\left\lbrace
\begin{split} 
& \frac{\pi k }{2 p} + \frac{\pi k^3 }{8 p^3} + \mathcal{O} \left( \frac{k^5}{p^5} \right) \, , \quad  p \gg k \\ 
& \frac{\pi }{2 } + \frac{ \pi p^2}{8 k^2} + \mathcal{O}\left( \frac{p^4}{k^4} \right) \, , \quad k \gg p  \, .
\end{split}
\right.
\end{equation}
In a simplified way, we can write these functions as in Eq.~\eqref{fSerie2} and Eq.~\eqref{gSerie2},
where the coefficients are obtained from Eq.~\eqref{fSerie} and Eq.~\eqref{gSerie}, respectively.



\begin{thebibliography}{99}

	\bibitem{Burden} C. J. Burden, J. Praschifka, and C. D. Roberts, \textit{Photon polarization tensor and gauge dependence in three-dimensional quantum electrodynamics}, Phys. Rev. D 46, 2695 (1992); 
G. Grignani, G. Semenoff, and P. Sodano, \textit{Confinement-deconfinement transition in three-dimensional QED}, Phys. Rev. D 53, 7157 (1996); 
P. Maris, \textit{Confinement and complex singularities in three-dimensional QED}, Phys. Rev. D 52, 6087 (1995).

	\bibitem{Maris} P. Maris,  \textit{Analytic structure of the full fermion propagator in quenched and unquenched QED}, Phys. Rev. D 50, 4189 (1994); T. Appelquist, M. J. Bowick, E. Cohler, and L. C. R. Wijewardhana, \textit{Chiral-Symmetry Breaking in 2+1 Dimensions}, Phys. Rev. Lett. 55, 1715 (1985).
	
	\bibitem{Prange} R. E. Prange and S. M. Girvin, \textit{The Quantum Hall Efect} (Springer, Berlin, 1987); F. Wilczek, \textit{Fractional Statistics and Anyon Superconductivity} (World Scientific, Singapore, 1990); R. B. Laughlin, \textit{Quantized Hall conductivity in tsto dimensions}, Phys. Rev. B 23, 5632 (1981).

	\bibitem{MarinoBook} E. C. Marino, \textit{Quantum Field Theory Approach to Condensed Matter Physics} (Cambridge University Press, Cambridge, England, 2017).

	\bibitem{Boulevard} Adriaan M. J. Schakel, \textit{Boulevard of broken symmetries: effective field theories of condensed matter}, (World Scientific, Singapore, 2008).
	
	\bibitem{Neto2009} A. H. Castro Neto, F. Guinea, N. M. R. Peres, K. S. Novoselov, and A. K. Geim, \textit{The electronic properties of graphene}, Rev. Mod. Phys. 81, 109 (2009).
	
	\bibitem{PRX} E. C. Marino, L. O. Nascimento, V. S. Alves, and C. M. Smith, \textit{Interaction Induced Quantum Valley Hall Effect in Graphene}, Phys. Rev. X 5, 011040 (2015).	
	
	\bibitem{Vozmediano1994} J. González, F. Guinea, and M. A. H. Vozmediano,  \textit{Non-Fermi liquid behavior of electrons in the half-filled honeycomb lattive (A renormalization group approach)}, Nucl. Phys. B 424, 595 (1994).
	
	\bibitem{Luis2020} L. Fernández, V. S. Alves, L. O. Nascimento, F. Peña, M. Gomes, and E. C. Marino, \textit{Renormalization of the band gap in 2D materials through the competition between electromagnetic and four-fermion interactions in large N expansion}, Phys. Rev. D 102, 016020 (2020).

	\bibitem{Marino1993} E. C. Marino, \textit{Quantum electrodynamics of particles on a plane and the Chern-Simons theory}, Nucl. Phys. B408, 551 (1993).	
	
	\bibitem{Amaral} R. L. P. G. do Amaral and E. C. Marinof, \textit{Canonical quantization of theories containing fractional powers of the d’Alembertian operator}, J. Phys. A 25, 5183 (1992).

	\bibitem{Marino2014} E. C. Marino, V. S. Alves, L. O. Nascimento, and C. M. Smith \textit{Unitarity of theories containing fractional powers of the d’Alembertian operator}, Phys. Rev. D 90, 105003 (2014).

	\bibitem{Ana} David Dudal, Ana Júlia Mizher, and Pablo Pais, \textit{Exact quantum scale invariance of three-dimensional reduced QED theories}, Phy. Rev. D 99, 045017 (2019).
	
	\bibitem{Amos} Matthew Heydeman, Christian B. Jepsen, Ziming Ji, and Amos Yarom, \textit{Renormalization and conformal invariance of non-local quantum electrodynamics}, J. High Energ. Phys. 2020, 7 (2020).

	\bibitem{FatorG} N. Menezes, V. S. Alves, E. C. Marino, L. Nascimento, L. O. Nascimento, and C. Morais Smith, \textit{Spin g-factor due to electronic interactions in graphene}, Phys. Rev. B 95, 245138 (2017).

	\bibitem{Excitons2018} E. C. Marino, L. O. Nascimento, V. S. Alves, N. Menezes, and C. M. Smith, \textit{Quantum-electrodynamical approach to the exciton spectrum in transition-metal dichalcogenides}, 2D Mater. 5, 041006 (2018).

	\bibitem{Alves2013} V. S. Alves, W. S. Elias, L. O. Nascimento, V. Juričić, and F. Peña, \textit{Chiral symmetry breaking in the pseudo-quantum electrodynamics}, Phys. Rev. D 87, 125002 (2013).

	\bibitem{Leandro} Leandro O. Nascimento, Van Sérgio Alves, Francisco Peña, C. Morais Smith, and E. C. Marino, \textit{Chiral-symmetry breaking in pseudoquantum electrodynamics at finite temperature}, Phys. Rev. D 92, 025018 (2015). 
 
	\bibitem{Luis2021} Luis Fernández, Reginaldo O. Corrêa, Jr., Van Sérgio Alves, Leandro O. Nascimento, and Francisco Peña. \textit{Dynamical mass generation in pseudoquantum electrodynamics with Gross-Neveu interaction at finite temperature}, Phy. Rev. D 103, 025018 (2021).

 	\bibitem{Menezes2016} N. Menezes, V. S. Alves, and C. Morais Smith, \textit{The influence of a weak magnetic field in the Renormalization-Group functions of (2+1)-dimensional Dirac systems}, Eur. Phys. J. B 89, 271 (2016).

	\bibitem{Danilo2017} J. D. Silva, A. N. Braga, W. P. Pires, V. S. Alves, D. T. Alves, and E. C. Marino,  \textit{Inhibition of the Fermi velocity renormalization in a graphene sheet by the presence of a conducting plate}, Nucl. Phys. B 920, 221 (2017).
	
	\bibitem{Cavity} Wagner P. Pires, Jeferson Danilo L. Silva, Alessandra N. Braga, Van Sérgio Alves, Danilo T. Alves, and E.C. Marino, \textit{Cavity effects on the Fermi velocity renormalization in a graphene sheet}, Nucl. Phys. B 932, 529 (2018).

	\bibitem{Miransky} E. V. Gorbar, V. P. Gusynin, and V. A. Miransky, \textit{Dynamical chiral symmetry breaking on a brane in reduced QED}, Phy. Rev. D, 64, 105028 (2001).
	
	\bibitem{Teber} S. Teber, \textit{Electromagnetic current correlations in reduced quantum electrodynamics}, Phy. Rev. D 86, 025005 (2012); S. Teber, \textit{Two-loop fermion self-energy and propagator in reduced QED3;2}, Phy. Rev. D 89, 067702 (2014); A. V. Kotikov and S. Teber, \textit{Two-loop fermion self-energy in reduced quantum electrodynamics and application to the ultrarelativistic limit of graphene}, Phy. Rev. D 89, 065038 (2014).

	\bibitem{Pedrelli} Danilo C. Pedrelli, Danilo T. Alves, and Van Sérgio Alves, \textit{Two-loop photon self-energy in pseudoquantum electrodynamics
in the presence of a conducting surface}, Phys. Rev. D 102, 125032 (2020).

	\bibitem{Ashcroft} N. W. Ashcroft and N. D. Mermin, \textit{Solid state physics} (Saunders College, Philadelphia, USA, 1976).

	\bibitem{Fig4_WSe2} P. V. Nguyen, N. C. Teutsch, N. P. Wilson, J. Kahn, X. Xia, A. J. Graham, V. Kandyba, A. Giampietri, A. Barinov, G. C. Constantinescu, N. Yeung, N. D. M. Hine, X. Xu, D. H. Cobden, and N. R. Wilson, \textit{Visualizing electrostatic gating effects in two-dimensional heterostructures}, Nature (London) 572, 220 (2019).
	
	\bibitem{Fig4_MoS2} F. Liu, M. E. Ziffer, K. R. Hansen, J. Wang, and X. Zhu, \textit{Direct Determination of Band-Gap Renormalization in the Photoexcited Monolayer MoS2}, Phys. Rev. Lett. 122, 246803 (2019).
			
	\bibitem{DasSarma2014} Edwin Barnes, E. H. Hwang, R. E. Throckmorton, and S. Das Sarma1, \textit{Effective field theory, three-loop perturbative expansion, and their experimental implications in graphene many-body effects}, Phy. Rev. B 89, 235431 (2014).	
	
	\bibitem{Matsubara} T. Matsubara, \textit{A New Approach to Quantum-Statistical Mechanics}, Progress of Theoretical Physics 14, 351 (1955).
	
	\bibitem{DasBookThermo} Ashok Das, \textit{Finite Temperature Field Theory} (World Scientific, Singapore, 1997).	

	\bibitem{Elias2011} D. C. Elias, R. V. Gorbachev, A. S. Mayorov, S. V. Morozov, A. A. Zhukov, P. Blake, L. A. Ponomarenko, I. V. Grigorieva, K. S. Novoselov, F. Guinea, and A. K. Geim, \textit{ Dirac cones reshaped by interaction effects in suspended graphene}, Nat. Phys. 7, 701 (2011).	
	
	\bibitem{Grads} I. S. Gradstheyn and I. M. Ryzhik, \textit{Table of Integrals, Series and Products, 7th ed.} (Academic Press, New York, 2007).
	
	\bibitem{Liu} Liu, HL., Yang, T., Chen, JH. et al. \textit{Temperature-dependent optical constants of monolayer MoS2, MoSe2, WS2, and WSe2: spectroscopic ellipsometry and first-principles calculations}, Sci. Rep. 10, 15282 (2020).
			
\end{thebibliography}
\end{document}